\documentclass[%
reprint,
 amsmath,amssymb,
 aps,
]{revtex4-1}

\usepackage{graphicx}
\usepackage{dcolumn}
\usepackage{bm}
\usepackage[mathlines]{lineno}
\usepackage[colorlinks=true,citecolor=blue,linkcolor=blue,anchorcolor=green, anchorcolor=green,urlcolor=blue]{hyperref}
\usepackage{mathrsfs}
\usepackage{xcolor}
\usepackage{float}
\usepackage{ulem}
\usepackage[utf8]{inputenc}

\begin{document}


\title{Innermost stable circular orbits of charged spinning test particles}

\author{Ming Zhang}
\email{mingzhang@mail.bnu.edu.cn}
\author{Wen-Biao Liu}
\email{wbliu@bnu.edu.cn}
\affiliation{Department of Physics, Beijing Normal University, Beijing, 100875, China}


\begin{abstract}
The effects of a paritcle's spin and electric charge on its angular momentum, energy and radius on the innermost stable circular orbit are investigated based on the particle's equations of motion in a background of the Kerr-Newmann spacetime. It is found that the particle's angular momentum and energy have monotonous relationships with not only its spin but also its charge; it is also discovered that the spinning particle's radius may change non-monotonously with its charge. Hence, our result remarkably indicates that particles owning identical spin but different charge may degenerate into a same last stable circular orbit.
\end{abstract}


\maketitle


\section{Introduction}
The innermost stable circular orbit (ISCO), as the name indicates, is the last stable circular orbit with a minimal radius for a particle revolving around the black hole \cite{wald1984general}. A particle will plunge into the black hole if its orbit radius is less than that of the ISCO (we denote this radius as $r_I$).

There are at least three reasons as we investigate the ISCO. Firstly, the spacetime geometry can be reflected by the ISCO of the particle. Secondly, the binary system consisting of a test particle and a black hole can be a source of gravitational waves \cite{Abbott:2016blz,Abbott:2016nmj}. What is more, the investigation of ISCO can give us a knowledge of the accretion disc and the related radiation spectrum \cite{Shakura:1972te,Page:1974he}.

For a massive particle without charge and spin, when it revolves around a Schwarzschild black hole, we know that the radius of its ISCO ($r_I$) is $6M$, with $M$ being the mass of the black hole \cite{chandrasekhar1985mathematical}. The radius of the ISCO for a Reissner-Nordstr\"{o}m (RN) black hole ranges from $4M$ (corresponding to an extreme black hole) to $6M$ (corresponding to the Schwarzschild limit) \cite{Pugliese:2010ps}.  When the particle moves around a Kerr black hole, the situation becomes a little more complicated. It was found that, in the backgroud of extreme Kerr black hole, $r_{I}=M$ for co-rotating orbits whereas $r_{I}=9M$ for counter-rotating orbits \cite{bardeen1972jm}. The investigation of the ISCO for the Kerr-Newmann (KN) black hole can be seen as a combination of the RN case and the Kerr case \cite{Liu:2017fjx}.

Other investigations about the ISCO can be seen in Refs. \cite{misner2017gravitation,Pugliese:2011xn,Pugliese:2011py,narayan2012observational,Tchekhovskoy:2011zx,Zahrani:2014rqa,Abdujabbarov:2009az,Akcay:2012ea,Asano:2016scy,Cabanac:2009yz,Campanelli:2006gf,Chakraborty:2013kza,Delsate:2015ina,Hadar:2011vj,Harms:2016ctx,Hod:2013vi,Hod:2014tpa,Isoyama:2014mja,Lukes-Gerakopoulos:2017vkj,Zaslavskii:2014mqa,Zahrani:2014rqa,Shiose:2014bqa,Hussain:2014cba,Takahashi:2008zh,Frolov:2010mi,Frolov:2011ea,Pugliese:2013zma,Zhang:2018omr} and many others, including the research for charged particles around KN black hole \cite{Schroven:2017jsp}. For convenience, we have listed typical cases in Table \ref{tbl:table1}, where $M, a, Q$ respectively stand for the mass, angular momentum and electric charge of the black hole, $j_I , e_I$ stand for the angular momentum and energy of the particle. 

\begin{table*}[!htbp]
\centering
\caption{ISCO parameters for an uncharged and spinless particle ($M=1$).}
\begin{tabular}{cccccc} 
  \hline\hline
$~~~~~~~~~a,~Q~~~~~~~~~$ &~~~~~~~~~Type of black hole~~~~~~~~~&~~~~~~~~~$r_I$~~~~~~~~~&~~~~~~~~~$j_I$~~~~~~~~~&~~~~~~~$e_I$~~~\\
\hline\hline 
$a=0,~Q=0$ & Schwarzschild & 6 & 2$\sqrt{3}$ & $\sqrt{\frac{8}{9}}$ \\
\hline
$a=0,~Q=1$ &  Extreme RN & 4 & $2\sqrt{2}$ & $\sqrt{\frac{27}{32}}$  \\
\hline
$a=0,~0<Q<1$ & RN & $(4,~6)$ & $(2\sqrt{2},~2\sqrt{3})$ & $\left(\sqrt{\frac{27}{32}},~\sqrt{\frac{8}{9}}\right)$  \\
\hline
$Q=0,~a=1$ & Extreme Kerr (co-rotating) & $1$ & $\frac{2}{\sqrt{3}}$ & $\frac{1}{\sqrt{3}}$ \\
\hline
$Q=0,~a=1$ & Extreme Kerr (counter-rotating) & $9$ & $-\frac{22}{3\sqrt{3}}$ & $\frac{5}{3\sqrt{3}}$ \\
\hline
$Q=0,~0<a<1$ & Kerr (co-rotating) &  $(1,~6)$  & $\left(\frac{2}{\sqrt{3}},~2\sqrt{3}\right)$ & $\left(\frac{1}{\sqrt{3}},~\sqrt{\frac{8}{9}}\right)$   \\
\hline
$Q=0,~0<a<1$ & Kerr (counter-rotating) &  $(6,~9)$  & $\left(-\frac{22}{3\sqrt{3}},~-2\sqrt{3}\right)$ & $\left(\sqrt{\frac{8}{9}},~\frac{5}{3\sqrt{3}}\right)$   \\
\hline
$a^2 +Q^2 \leqslant 1$ & (Extreme) KN (co-rotating) & $(1,~6)$ & $\left(\frac{2}{\sqrt{3}},~2\sqrt{3}\right)$ & $\left(\frac{1}{\sqrt{3}},~\sqrt{\frac{8}{9}}\right)$  \\
\hline
$a^2 +Q^2 \leqslant 1$ & (Extreme) KN (counter-rotating) &  $(4,~9)$ & $\left(-\frac{22}{3\sqrt{3}},~-2\sqrt{2}\right)$ & $\left(\sqrt{\frac{27}{32}},~\frac{5}{3\sqrt{3}}\right)$   &      \\
\hline
\hline
\end{tabular}
  \label{tbl:table1}
\end{table*}

In fact, a classical test body may contain spin. Effects of spin on the ISCO orbits are investigated for the Schwarzschild black hole \cite{Corinaldesi:1951pb,rasband1973black}, the Kerr black hole \cite{rasband1973black,Tod:1976ud} and the KN black hole \cite{Zhang:2017nhl}.

What will happen when the particle contains not only spin but also electric charge? In other words, what effects will the spin and charge of the particle take on the ISCO? In this paper, we will answer this question and show the interplay of the spin and the charge carried by the particle on the corresponding ISCO in the background of KN spacetime. We will write the equations of motion for a charged spinning test particle in KN black hole in Sec. \ref{eos}. The effects of the ISCO parameters $j, e, r_I$ of the charged and spinning particle revolving around the KN black hole will be shown in Sec. \ref{iscounless}. Our conclusion will be given in Sec. \ref{con}.

\section{The equations of motion of a charged spinning particle in KN spacetime}\label{eos}
The metric of the KN spacetime is
\begin{equation}\label{metric}
\begin{aligned}
ds^2=&-\left(1-\frac{2 Mr-Q^2}{\Sigma }\right)dt^2 +\frac{\Sigma }{\Delta}dr^2+ \Sigma d\theta^2\\&+\frac{\left(a^2+r^2\right)^2-a^2 \Delta  \sin ^2 \theta }{\Sigma }\sin ^2\theta d\phi^2\\&-\frac{2 a \left(2Mr-Q^{2}\right)}{\Sigma }\sin ^2\theta dt d\phi, 
\end{aligned}
\end{equation}
where 
\begin{equation}
\Sigma=r^2+a^2 \cos ^2\theta,~\Delta =a^2-2 Mr+Q^2+r^2\nonumber.
\end{equation}
The gauge field reads 
\begin{equation}
F=dA,~A_a=-\frac{Qr}{\Sigma}\left(dt-a \text{sin}^2 \theta d\phi\right).
\end{equation}
The outer event horizon locates at
\begin{equation}
r_+=M+\sqrt{M^2-a^2-Q^2}.
\end{equation}
After choosing the tetrad
\begin{equation}\label{tetrad}
e_a^{(0)}=\sqrt{\frac{\Delta }{\Sigma }} \left(dt-a \sin^{2}\theta d\phi \right),~e_a^{(1)}=\sqrt{\frac{\Sigma }{\Delta }}dr,\nonumber
\end{equation}
\begin{equation}
e_a^{(2)}= \sqrt{\Sigma }d\theta,~e_a^{(3)}=\frac{\sin \theta}{\sqrt{\Sigma }} \left[-a dt+\left(a^2+r^2\right)d\phi\right],\nonumber
\end{equation}
the metric Eq. (\ref{metric}) can be rewritten as
\begin{equation}
ds^2=\eta_{(i)(j)}e_{a}^{(i)}e_{b}^{(j)}.
\end{equation}

Actually, considering spin and electric charge endowed to an astronomical test particle, the particle (which is supposed to be a charged top, i.e., the charge of spinning body is gathered on a ceter point and the magnetic moment of the particle can be ignored \cite{Hojman:1976kn}) does not move along the geodesic. Instead, its motion should be described by the Mathisson-Papapetrou-Dixon (MPD) equations \cite{Hojman:1976kn}
\begin{equation}\label{mp1}
\frac{D P^a}{D \tau }=-\frac{1}{2}R^a{}_{bcd}v^{b} S^{cd} -\mathcal{Q}F^{a}_{b } v^{b},
\end{equation}
\begin{equation}\label{mp2}
\frac{D S^{\text{ab}}}{D \tau }=2P^{[a}v^{b]}.
\end{equation}
In Eqs. (\ref{mp1}), (\ref{mp2}), $P^{a}$ stands for the particle's 4-momentum, $\tau$ is the affine parameter (proper time),  $v^{a}$ is the 4-velocity, $S^{ab}$ represents the spin tensor, $\mathcal{Q}$ is the charge of the particle, $F^{ab}$ denotes the electromagnetic field tensor of the spacetime.

A normalized 4-momentum $u^a$ can be defined as
\begin{equation}
u^a\equiv \frac{P^a}{m}.
\end{equation}
Three other supplementary conditions are
\begin{equation}
S^{ab}S_{ab}=2 S^2,~S^{ab} P_b=0,~u^{a}v_{a}=-1.\nonumber
\end{equation}
The first one signifies that the magnitude of the spin $S$ is conserved; the second one ensures the conservation of the dynamical particle's mass $m$ \cite{Saijo:1998mn}; the last one is used to normalize the proper time $\tau$ \cite{Saijo:1998mn}. 

Then, one can obtain \cite{Hojman:1976kn}
\begin{equation}\label{relationuv}
v^a-u^a=\frac{2 S^{ab} u^c \left(2 \mathcal{Q}F_{bc}+R_{bcde} S^{de}\right)}{S^{bc} \left(2 \mathcal{Q}F_{bc}+R_{bcde} S^{de}\right)+4 m^2}.
\end{equation}
We can see that the 4-velocity is no longer parallel to the 4-momentum due to the emergence of the particle's spin.

There are two conserved quantities $\tilde{E}$ and $\tilde{J}$, which correspond to the timelike Killing vector $\left(\frac{\partial }{\partial t}\right)^a$ and axial Killing vector $\left(\frac{\partial }{\partial \phi }\right)^a$ respectively and can be written as \cite{Hojman:1976kn}
\begin{equation}
e\equiv\frac{\tilde{E}}{m}=\frac{1}{2 m}S^{ab} \nabla _b\xi _a-\xi _a u^a-q\phi,
\end{equation}
\begin{equation}
j\equiv\frac{\tilde{J}}{m}=-\frac{1}{2 m}S^{ab} \nabla _b\phi _a+u^a \phi _a+qh,
\end{equation}
where 
\begin{equation}
q=\frac{\mathcal{Q}}{m},~\phi =-\frac{Q r}{\Sigma },~h=\frac{Q a r }{\Sigma}\sin^{2}\theta.\nonumber
\end{equation}

Specifically, a spin vector $s^{a}$ can be choosed as
 \begin{equation}
s^a=-\frac{1}{2m}\varepsilon^{(a)} {}_{(b) (c) (d)}u^{(b)} S^{(c) (d)},
\end{equation}
$\varepsilon _{(a) (b) (c) (d)}$ here is a completely antisymmetric tensor and $\varepsilon _{(1)(2)(3)(4)}=1$.
We set the only nonvanishing component of $s^{(a)}$ as
\begin{equation}
s^{(2)}=-s,
\end{equation}
where $s$ represents the magnitude of the spin and $s>0$ manifests that the direction of the spin is parallel to that of the rotating black hole. As a result, the nonvanishing tetrad components of the spin tensor can be obtained as
\begin{equation}
S^{(0)(1)}=-m s u^{(3)},~S^{(0)(3)}=m s u^{(1)},~S^{(1)(3)}=m s u^{(0)}.
\end{equation}
As the stable circular orbits of the pariticle around a rotating black hole locates in the equatorial plane, we will set $\theta=\pi/2$ hereafter. The conserved quantities can be further expressed as
\begin{eqnarray}\begin{aligned}
e=&\frac{\sqrt{\Delta}}{r}u^{(0)}+\frac{a r^2+M r s-sQ^2}{r^{3}}u^{(3)}+\frac{qQ}{r},\\
j=&\frac{\sqrt{\Delta}(a+s)}{r}u^{(0)}+\frac{\mathcal{P}_1}{r^{3}}u^{(3)}+\frac{aqQ}{r},
\end{aligned}\end{eqnarray}
where $\mathcal{P}_1 =r^4+ar^2 (a+s)+aMrs-asQ^2$.

The nonvanishing normalized momentum can be obtained as
\begin{equation}\begin{aligned}
u^{(0)}=&\frac{1}{\sqrt{\Delta}\mathcal{X}}\left[e r^5-q Q r^4+(ea+es-j)a r^3\right.\\&\left.+  (a e M-a q Q-j M)sr^{2}+ (j-a e)Q^2 s r\right],\\
\end{aligned}\end{equation}
\begin{equation}
u^{(3)}=\frac{r^3 (j-ea-es)+r^2 q Q s}{\mathcal{X}},
\end{equation}
\begin{equation}\label{ramo}
u^{(1)}=\sigma \sqrt{-1+(u^{(0)})^2  - (u^{(3)})^2}=\sigma \sqrt{O},
\end{equation}
where $\mathcal{X}=r^{4}-M r s^2+Q^2 s^2$, $\sigma=1$ corresponds to a radially outgoing particle and $\sigma=-1$ for a radially ingoing one. Applying Eq. (\ref{relationuv}), the 4-velocity can be obtained as
\begin{eqnarray}
v^{(0)}&=&\frac{ r^{4}+s^2 \left(Q^2-M r\right)}{\mathcal{P}_2}u^{(0)},\\
v^{(1)}&=&\frac{r^{4}+s^2 \left(Q^2-M r\right)}{\mathcal{P}_2}u^{(1)},\label{proone}\\
v^{(3)}&=&\frac{r^4+2 M r s^2-3 Q^2 s^2}{\mathcal{P}_2}u^{(3)}-\frac{q Q r^2 s}{\mathcal{P}_2},
\end{eqnarray}
where
\begin{equation}\begin{aligned}
\mathcal{P}_2=&\left(4 s^2 Q^2-3M r\right)(u^{(3)})^{2}+qQ r^2 s u^{(3)}/m \\&+s^2 \left(Q^2-M r\right)+r^4.\nonumber
\end{aligned}\end{equation}

The 4-velocity of the particle can be expressed as
\begin{equation}
v^a=\frac{dt}{d\tau}\left(\frac{\partial }{\partial t}\right)^a+\frac{dr}{d\tau}\left(\frac{\partial }{\partial r}\right)^a+\frac{d\phi}{d\tau}\left(\frac{\partial}{\partial \phi }\right)^a.
\end{equation}
According to the tetrad transformation relations Eq. (\ref{tetrad}), one can obtain
\begin{eqnarray}
v^{(0)}&=&\sqrt{\frac{\Delta }{\Sigma }} \left(\frac{dt}{d\tau}-a\sin ^2\theta\frac{d\phi }{\text{d$\tau $}}\right)\label{velocityone},\\
v^{(1)}&=& \sqrt{\frac{\Sigma }{\Delta }}\frac{dr}{d\tau},\\
v^{(3)}&=&\frac{\sin (\theta ) }{\sqrt{\Sigma }}\left[\left(a^2+r^2\right)\frac{d\phi}{d\tau }-\frac{a dt}{d\tau }\right].\label{velocitytwo}
\end{eqnarray}
Next up, the equations of motion for a charged spinning particle in the background of KN spacetime can be obtained as
\begin{equation}\label{eos1}
\frac{dt}{d\tau}=\frac{\mathcal{X} \left(a^2 \mathcal{P}_4 \mathcal{X}+a \Delta  r\mathcal{P}_5 +\mathcal{P}_4   r^2 \mathcal{X} \right)}{\sqrt{\Delta}  \left[ -3 M {\mathcal{P}_3}^2  s^2 r^{5}+ \left(4 {\mathcal{P}_3}^2 Q^2 s^2+{\mathcal{X}}^2\right)r^{4}+\mathcal{P}_6 \right]},
\end{equation}
\begin{equation}\label{rtau}
\frac{dr}{d\tau}=\sqrt{\frac{\Delta}{\Sigma}}v^{(1)},
\end{equation}
\begin{equation}\label{phitau}
\frac{d\phi}{d\tau}=\frac{1}{a \sin\theta^2}\left(\frac{dt}{d\tau}-\sqrt{\frac{\Sigma}{\Delta}}v^{(0)}\right),
\end{equation}
where
\begin{eqnarray}
\mathcal{P}_3&=&r \left[j-e (a+s)\right]+q Q s\nonumber,\\
\mathcal{P}_4 &=&a^2 e r^2-a \left[e s \left(Q^2-Mr+r^{2}\right)+r (j r+q Q s)\right]\nonumber\\&&+r^3 (e r-q Q)+j s \left(Q^2-M r\right)\nonumber,\\
\mathcal{P}_5&=&  P \left(2 M r s^2-3 Q^2 s^2+r^4\right)-q Q R s/m,\nonumber\\
\mathcal{P}_6 &=&-M r X^2 s^2+Q^2 X^2 s^2+P q Q r^4 X s/m.\nonumber
\end{eqnarray}
As concrete expressions for Eqs. (\ref{rtau}), (\ref{phitau}) can be obtained by simple algebraic calculation, we do not show them explicitly here.

It should be noticed that the motion of the particle must obey the forward-in-time condition $dt/d\tau\geqslant 0$ as well as the the timelike 4-velocity condition $v^{\mu}v_{\mu}<0$. We should also keep in mind that $s\ll M$ \cite{Wald:1972sz} and $q\ll M$.

 \begin{figure*}[!htbp] 
   \centering
   \includegraphics[width=5in]{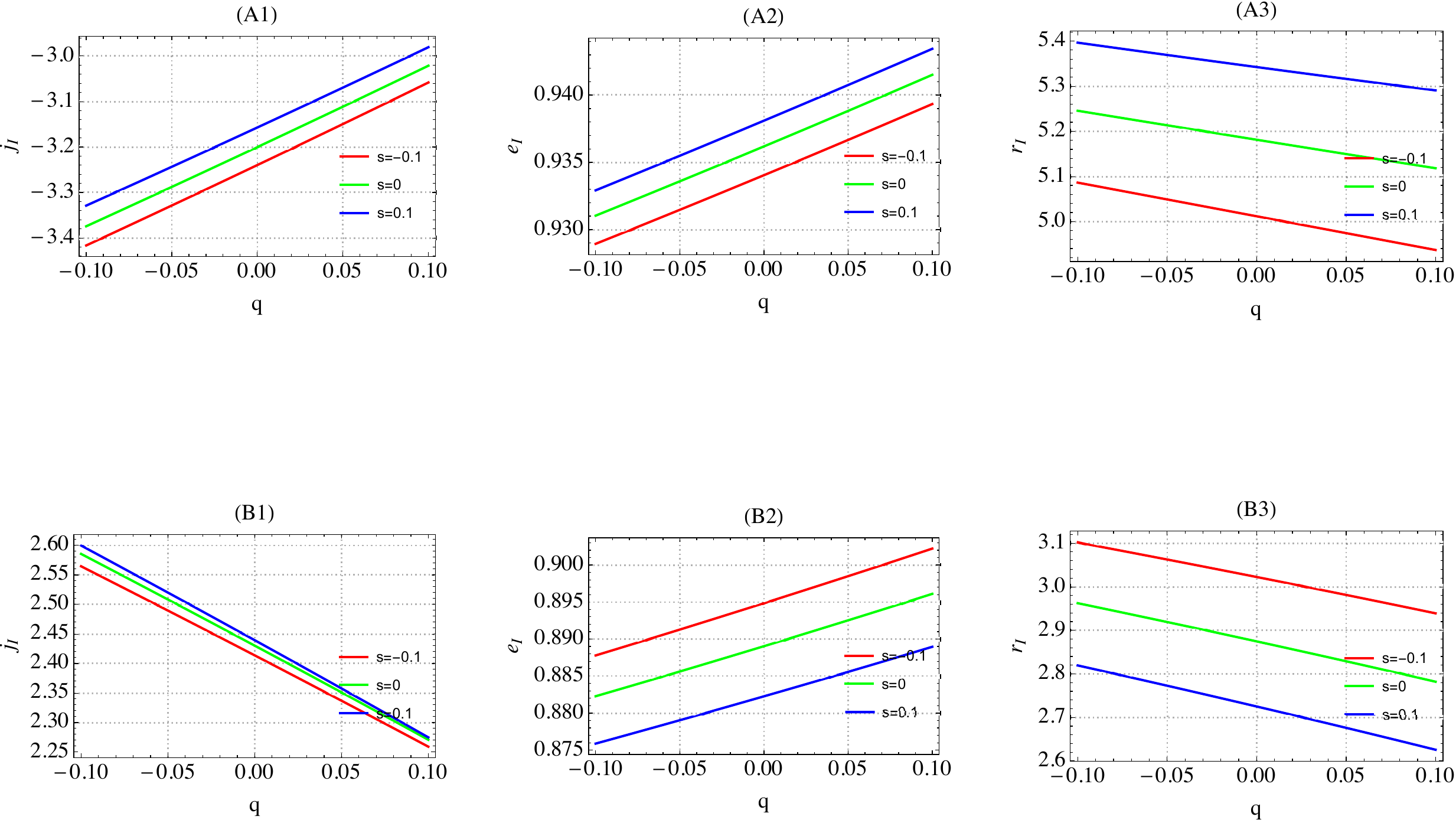}
   \caption{The diagrams show effects of spin and charge on the ISCO parameters $j_I, e_I, r_I$ for $M=1, a=1/4, Q=\sqrt{15}/4$. Among them, (A1)-(A3) are ISCO parameters for counter-rotating orbits and (B1)-(B3) are ISCO parameters for co-rotating orbits.}
   \label{ef1}
\end{figure*}

\section{The ISCO of a charged spinning particle around the KN black hole}\label{iscounless}
The trajectory of the test particle is stable iff
\begin{equation}\label{conone}
\frac{dr}{d\tau}=0,
\end{equation}
\begin{equation}\label{contwo}
\frac{d^2 r}{d{\tau}^2}=0.
\end{equation}
Eq. (\ref{conone}) ensures that the particle does not have radial velocity while Eq. (\ref{contwo}) gives a restriction that there is no radial acceleration for the particle.

The effective potential can be defined as the minimum allowable value of the energy for the pariticle at radius $r$. As the radial velocity $dr/d\tau$ are proportional to the radial component of the four-momentum $u^{(1)}$ [see Eqs. (\ref{proone}), (\ref{rtau})], we can use $u^{(1)}$ to define the effective potential. According to Eq. (\ref{ramo}), we can obtain the square of the normalized four-momentum $u^{(1)}$ as 
\begin{equation}\label{squramo}
(u^{(1)})^2=\frac{\alpha e^2 +\beta e + \gamma}{\mathcal{X}^2 \Delta},
\end{equation}
where
\begin{equation}\begin{aligned}
\alpha = &2 a^3 r^4 s \left(M r-Q^2\right)+2 a r^6 s \left(3 M r-2 Q^2\right)\\&+r^6 \left(2 M r s^2-Q^2 s^2+r^4-r^2 s^2\right)\\&+a^2 r^2 \left(Q^4 s^2-Q^2 r \left(2 M s^2+r^3+2 r s^2\right)\right)\\&+a^2 r^4 \left(M^2 s^2+2 M r \left(r^2+s^2\right)+r^4\right),
\end{aligned}\end{equation}
\begin{equation}\begin{aligned}
\beta = & \beta_s + q\left[-2 a^2 M Q r^4 s^2+2 a^2 Q^3 r^3 s^2-2 Q r^9\right.\\&\left. +r^7 \left(-2 a^2 Q-2 a Q s+2 Q s^2\right)\right.\\&\left. +r^6 \left(-6 a M Q s-4 M Q s^2\right)+r^5 \left(4 a Q^3 s+2 Q^3 s^2\right)\right],
\end{aligned}\end{equation}
\begin{equation}\begin{aligned}
\gamma = &\gamma_s +j r^3 Q q\left[-2 r^2 s \left(r^2-3 M r+2 Q^2\right)\right.\\&\left.+a \left(2 M r s^2+Q^2 s^2+2 r^4\right)\right]\\&+Q^2 r^4 q^2 \left[r^2 s (2 a-s)+2 M r s^2-Q^2 s^2+r^4\right]
\end{aligned}\end{equation}
with 
\begin{equation}\begin{aligned}
\beta_s = & 4 a^2 j r^4 s \left(Q^2-M r\right)+2 j r^6 s \left(r (r-3 M)+2 Q^2\right)\\&+2 a j Q^2 r^3 \left(2 M s^2+r^3+r s^2\right)\\&-2 a j r^2 \left(M r^2 \left(M s^2+2 r^3+r s^2\right)+Q^4 s^2\right),
\end{aligned}\end{equation}
\begin{equation}\begin{aligned}
\gamma_s = &-\Delta  \left(-M r s^2+Q^2 s^2+r^4\right)^2\\&+j^2 r^2 \left[Q^4 s^2-Q^2 r \left(2 a r s+2 M s^2+r^3\right)\right]\\&+j^2 r^4 \left[2 M \left(a r s+r^3\right)+M^2 s^2-r^4\right].
\end{aligned}\end{equation}
Then the effective potential can be defined as
\begin{equation}\label{veff}
V\equiv\frac{-\beta+\sqrt{\beta ^2-4 \alpha  \gamma }}{2 \alpha }.
\end{equation}

 \begin{figure*}[!htbp] 
   \centering
   \includegraphics[width=7in]{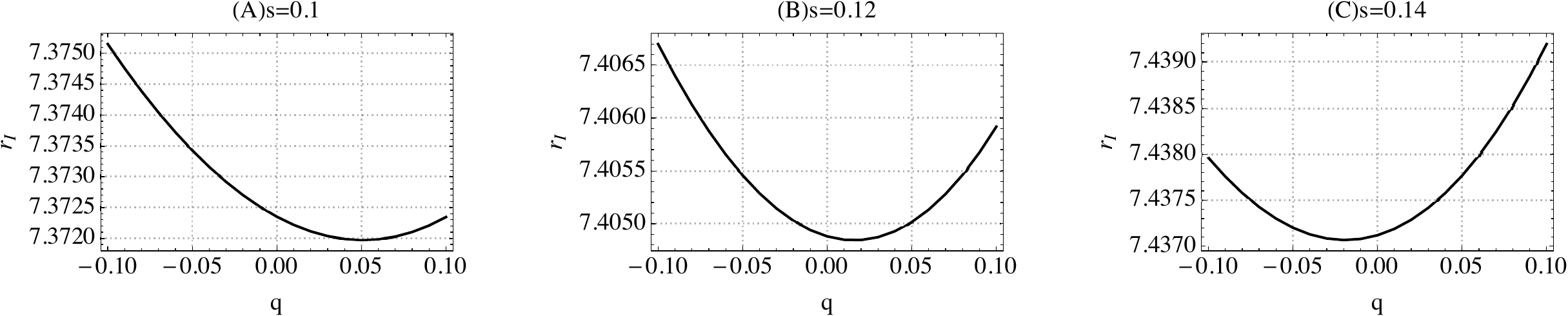}
   \caption{The diagrams show effects of spin and charge on the counter-rotating ISCO radius $r_I$ for $M=1, a=1/2, Q=1/2$.}
   \label{ef2}
\end{figure*}

The ISCO locates at a point where the maximum and minimum of the effective potential merge, which means
\begin{equation}\label{conthree}
\frac{d^2 V}{dr^2}=0.
\end{equation}
These three conditions Eqs. (\ref{conone}), (\ref{contwo}), (\ref{conthree}) can be used to calculate the parameters $r_I,~j_I ,~e_I$ for ISCO of the test particle. Generallly, we can only obtain the numerical results for the set of nonlinear equations.

Here we will put our emphasis on elaborating the interplay of spin and charge endowed to the particle. As shown, we have plotted the diagrams reflecting the effects of spin and charge in Fig. \ref{ef1}. 

Firstly, let us see the effects of charge and spin hold by the particle on the conserved angular momentum. From the diagram (A1) in Fig. \ref{ef1}, we can see that the conserved angular momentum increases with the increasing electric charge and spin for the particles revolving on the counter-rotating orbit. On the other hand, from the diagram (B1) in Fig. \ref{ef1}, it is obvious that the conserved angular momentum of the particle decreases with the increasing charge taken by the particle while it increases with the spin for the co-rotating orbit.

Secondly, we can know from (A2) in Fig. \ref{ef1} that the conserved energy of the counter-rotating particle increases with the charge and spin hold by the particle. However, for the co-rotating particle as shown in (B2) of Fig. \ref{ef1}, its conserved energy increases with the increasing positive charge but decreases with the spin.

We can see that the effects of the spin and charge taken by the particle on the conserved angular momentum and conserved energy are monotonous. Though we only show results of limited values of spin and charge for the particle in Fig. \ref{ef1}, we have checked our conclusion for all possible values of spin and charge taken by the particle in background of other extreme and non-extreme KN black hole where no counterexample has been found. 

Now we will focus our discussion on the ISCO radius. Though the effects of spin and charge are monotonous in most cases, as shown in Fig. \ref{ef1}; we can see counterexamples, as shown in Fig. \ref{ef2}. Remarkably, as Fig. \ref{ef2} shows, two particles with the same spin but different charge can occupy a same ISCO orbit. In other words, the radii of the ISCO orbit become degenerated for particles with the same spin but different charges in certain spacetime background. At the first sight, it seems incredible; however, this is a result stemming from the interplay of the spin and charge taken by the particle. In fact, one can find from the effective potential Eq. (\ref{veff}) that there exists terms relating to $qs,~qs^2,~q^2 s,~q^2 s^2$.

\section{Conclusion and Discussion}\label{con}
In this paper, we have done analyses for the effects of spin and electric charge on the ISCO parameters 
$j_I, e_I, r_I$, representing the conserved angular momentum, conserved energy, and innermost radius respectively. To that end, we have written the equations of motion for a charged spinning particle in the KN spacetime background at first, and then have numerically calculated the ISCO parameters according to the effective potential for the charged spinning particles. 

Our results show that the spin and charge hold by the particle affect the conserved angular momentum $j_I$ and the conserved energy $e_I$ monotonously. To be specific, we have shown that the angular momentum and energy of the particle on the counter-rotating ISCO increase with the spin and charge hold by the particle. For the particle on the co-rotating ISCO, the situation becomes a little more complicated: the angular momentum decreases with the increasing charge and spin of the particle; the energy increases with the decreasing spin and increasing charge. These results are in consistency with those in Ref. \cite{Schroven:2017jsp} where the effect of the particle's charge is investigated and in Ref. \cite{Zhang:2017nhl} where the effect of the particle's spin is researched.

Our results also show that the spinning particle's radius may change non-monotonously with its charge. As a result, we can see that particles owning identical spin but different charge may degenerate into one ISCO. Thus, we reveal the coupling effect of the charge and spin taken by the particle on the ISCO radius.

Investigations for pariticle motion in the vicinity of the black hole can provide valuable references for the study of astrophysical events \cite{Narayan:2011eb} and high energy events \cite{Zhang:2018gpn} relating to the black hole. Our result about the degeneration of the particle orbits may provide useful theoretical prediction of the observation of the electromagnetic waves as well as the gravitational waves. Thinking of the spacetime background in Fig. \ref{ef2}, one can envisage that two particles endowed with identical mass and spin (say, $s=0.12$) but opposite electrical charge (one can set $q_1 =0.05, q_2=-0.02$) move at the same ISCO in the same directions and have an elastic collision with each other, resulting in other two particles with the same spin but less charge (as the charge is neutralized), then both of these two particles will have a smaller ISCO orbit after the instantaneous event. We suspect that there must be astrophysically observable phenomenon corresponding to this interesting collisional event.

There is another important fact relating to the degeneration of the ISCO for particles with identical spin but different charge, that is, particles on the same ISCO have different conserved angular momenta and conserved energies. According to the equations of motion Eqs. (\ref{eos1}) and (\ref{phitau}) for the spinning charge particle, the angular velocities $\Omega$ (also called as Kepler frequencies) of the particles observed at infinity, which can be defined as
\begin{equation}
\Omega\equiv\frac{d\phi}{dt}=\frac{d\phi/d\tau}{dt/d\tau},
\end{equation}
will be different. So classically we can predict that two particles with identical spin but different charge on the same ISCO will release different electromagnetic \cite{Castineiras:2005ww} and gravitational radiations \cite{Misner:1972kx,Misner:1972jf} before they collide with each other. What's more, on the one hand, one must notice that semiclassically we should analyze this scenario in the context of quantum field theory in KN spacetime \cite{Poisson:1993vp,Cutler:1993vq,Bernar:2017kug}, on the other hand, as gravitational radiations from a spinning particle around a rotating black hole has been calculated in Teukolsky-Sasaki-Nakamura formalism \cite{Harms:2016ctx,Saijo:1998mn,Han:2010tp,Harms:2015ixa}, it must be interesting to extend their studies to the charged spinning particle case.

\section*{Acknowledgements}
This work is supported by the National Natural Science Foundation of China (Grant No. 11235003).

\bibliographystyle{apsrev4-1}
\bibliography{Notes}

\begin{thebibliography}{54}%
\makeatletter
\providecommand \@ifxundefined [1]{%
 \@ifx{#1\undefined}
}%
\providecommand \@ifnum [1]{%
 \ifnum #1\expandafter \@firstoftwo
 \else \expandafter \@secondoftwo
 \fi
}%
\providecommand \@ifx [1]{%
 \ifx #1\expandafter \@firstoftwo
 \else \expandafter \@secondoftwo
 \fi
}%
\providecommand \natexlab [1]{#1}%
\providecommand \enquote  [1]{``#1''}%
\providecommand \bibnamefont  [1]{#1}%
\providecommand \bibfnamefont [1]{#1}%
\providecommand \citenamefont [1]{#1}%
\providecommand \href@noop [0]{\@secondoftwo}%
\providecommand \href [0]{\begingroup \@sanitize@url \@href}%
\providecommand \@href[1]{\@@startlink{#1}\@@href}%
\providecommand \@@href[1]{\endgroup#1\@@endlink}%
\providecommand \@sanitize@url [0]{\catcode `\\12\catcode `\$12\catcode
  `\&12\catcode `\#12\catcode `\^12\catcode `\_12\catcode `\%12\relax}%
\providecommand \@@startlink[1]{}%
\providecommand \@@endlink[0]{}%
\providecommand \url  [0]{\begingroup\@sanitize@url \@url }%
\providecommand \@url [1]{\endgroup\@href {#1}{\urlprefix }}%
\providecommand \urlprefix  [0]{URL }%
\providecommand \Eprint [0]{\href }%
\providecommand \doibase [0]{http://dx.doi.org/}%
\providecommand \selectlanguage [0]{\@gobble}%
\providecommand \bibinfo  [0]{\@secondoftwo}%
\providecommand \bibfield  [0]{\@secondoftwo}%
\providecommand \translation [1]{[#1]}%
\providecommand \BibitemOpen [0]{}%
\providecommand \bibitemStop [0]{}%
\providecommand \bibitemNoStop [0]{.\EOS\space}%
\providecommand \EOS [0]{\spacefactor3000\relax}%
\providecommand \BibitemShut  [1]{\csname bibitem#1\endcsname}%
\let\auto@bib@innerbib\@empty
\bibitem [{\citenamefont {Wald}(1984)}]{wald1984general}%
  \BibitemOpen
  \bibfield  {author} {\bibinfo {author} {\bibfnamefont {R.}~\bibnamefont
  {Wald}},\ }\href@noop {} {\enquote {\bibinfo {title} {General relativity
  (chicage, il)},}\ } (\bibinfo {year} {1984})\BibitemShut {NoStop}%
\bibitem [{\citenamefont {Abbott}\ \emph
  {et~al.}(2016{\natexlab{a}})\citenamefont {Abbott} \emph
  {et~al.}}]{Abbott:2016blz}%
  \BibitemOpen
  \bibfield  {author} {\bibinfo {author} {\bibfnamefont {B.~P.}\ \bibnamefont
  {Abbott}} \emph {et~al.} (\bibinfo {collaboration} {Virgo, LIGO
  Scientific}),\ }\href {\doibase 10.1103/PhysRevLett.116.061102} {\bibfield
  {journal} {\bibinfo  {journal} {Phys. Rev. Lett.}\ }\textbf {\bibinfo
  {volume} {116}},\ \bibinfo {pages} {061102} (\bibinfo {year}
  {2016}{\natexlab{a}})},\ \Eprint {http://arxiv.org/abs/1602.03837}
  {arXiv:1602.03837 [gr-qc]} \BibitemShut {NoStop}%
\bibitem [{\citenamefont {Abbott}\ \emph
  {et~al.}(2016{\natexlab{b}})\citenamefont {Abbott} \emph
  {et~al.}}]{Abbott:2016nmj}%
  \BibitemOpen
  \bibfield  {author} {\bibinfo {author} {\bibfnamefont {B.~P.}\ \bibnamefont
  {Abbott}} \emph {et~al.} (\bibinfo {collaboration} {LIGO Scientific,
  Virgo}),\ }\href {\doibase 10.1103/PhysRevLett.116.241103} {\bibfield
  {journal} {\bibinfo  {journal} {Phys. Rev. Lett.}\ }\textbf {\bibinfo
  {volume} {116}},\ \bibinfo {pages} {241103} (\bibinfo {year}
  {2016}{\natexlab{b}})},\ \Eprint {http://arxiv.org/abs/1606.04855}
  {arXiv:1606.04855 [gr-qc]} \BibitemShut {NoStop}%
\bibitem [{\citenamefont {Shakura}\ and\ \citenamefont
  {Sunyaev}(1973)}]{Shakura:1972te}%
  \BibitemOpen
  \bibfield  {author} {\bibinfo {author} {\bibfnamefont {N.~I.}\ \bibnamefont
  {Shakura}}\ and\ \bibinfo {author} {\bibfnamefont {R.~A.}\ \bibnamefont
  {Sunyaev}},\ }\href@noop {} {\bibfield  {journal} {\bibinfo  {journal}
  {Astron. Astrophys.}\ }\textbf {\bibinfo {volume} {24}},\ \bibinfo {pages}
  {337} (\bibinfo {year} {1973})}\BibitemShut {NoStop}%
\bibitem [{\citenamefont {Page}\ and\ \citenamefont
  {Thorne}(1974)}]{Page:1974he}%
  \BibitemOpen
  \bibfield  {author} {\bibinfo {author} {\bibfnamefont {D.~N.}\ \bibnamefont
  {Page}}\ and\ \bibinfo {author} {\bibfnamefont {K.~S.}\ \bibnamefont
  {Thorne}},\ }\href {\doibase 10.1086/152990} {\bibfield  {journal} {\bibinfo
  {journal} {Astrophys. J.}\ }\textbf {\bibinfo {volume} {191}},\ \bibinfo
  {pages} {499} (\bibinfo {year} {1974})}\BibitemShut {NoStop}%
\bibitem [{\citenamefont {Chandrasekhar}\ and\ \citenamefont
  {Thorne}(1985)}]{chandrasekhar1985mathematical}%
  \BibitemOpen
  \bibfield  {author} {\bibinfo {author} {\bibfnamefont {S.}~\bibnamefont
  {Chandrasekhar}}\ and\ \bibinfo {author} {\bibfnamefont {K.~S.}\ \bibnamefont
  {Thorne}},\ }\href@noop {} {\enquote {\bibinfo {title} {The mathematical
  theory of black holes},}\ } (\bibinfo {year} {1985})\BibitemShut {NoStop}%
\bibitem [{\citenamefont {Pugliese}\ \emph
  {et~al.}(2011{\natexlab{a}})\citenamefont {Pugliese}, \citenamefont
  {Quevedo},\ and\ \citenamefont {Ruffini}}]{Pugliese:2010ps}%
  \BibitemOpen
  \bibfield  {author} {\bibinfo {author} {\bibfnamefont {D.}~\bibnamefont
  {Pugliese}}, \bibinfo {author} {\bibfnamefont {H.}~\bibnamefont {Quevedo}}, \
  and\ \bibinfo {author} {\bibfnamefont {R.}~\bibnamefont {Ruffini}},\ }\href
  {\doibase 10.1103/PhysRevD.83.024021} {\bibfield  {journal} {\bibinfo
  {journal} {Phys. Rev.}\ }\textbf {\bibinfo {volume} {D83}},\ \bibinfo {pages}
  {024021} (\bibinfo {year} {2011}{\natexlab{a}})},\ \Eprint
  {http://arxiv.org/abs/1012.5411} {arXiv:1012.5411 [astro-ph.HE]} \BibitemShut
  {NoStop}%
\bibitem [{\citenamefont {Bardeen}(1972)}]{bardeen1972jm}%
  \BibitemOpen
  \bibfield  {author} {\bibinfo {author} {\bibfnamefont {J.~M.}\ \bibnamefont
  {Bardeen}},\ }\href@noop {} {\bibfield  {journal} {\bibinfo  {journal}
  {Astrophys. J.}\ }\textbf {\bibinfo {volume} {178}},\ \bibinfo {pages} {347}
  (\bibinfo {year} {1972})}\BibitemShut {NoStop}%
\bibitem [{\citenamefont {Liu}\ \emph {et~al.}(2017)\citenamefont {Liu},
  \citenamefont {Lee},\ and\ \citenamefont {Lin}}]{Liu:2017fjx}%
  \BibitemOpen
  \bibfield  {author} {\bibinfo {author} {\bibfnamefont {C.-Y.}\ \bibnamefont
  {Liu}}, \bibinfo {author} {\bibfnamefont {D.-S.}\ \bibnamefont {Lee}}, \ and\
  \bibinfo {author} {\bibfnamefont {C.-Y.}\ \bibnamefont {Lin}},\ }\href
  {\doibase 10.1088/1361-6382/aa903b} {\bibfield  {journal} {\bibinfo
  {journal} {Class. Quant. Grav.}\ }\textbf {\bibinfo {volume} {34}},\ \bibinfo
  {pages} {235008} (\bibinfo {year} {2017})},\ \Eprint
  {http://arxiv.org/abs/1706.05466} {arXiv:1706.05466 [gr-qc]} \BibitemShut
  {NoStop}%
\bibitem [{\citenamefont {Misner}\ \emph {et~al.}(2017)\citenamefont {Misner},
  \citenamefont {Thorne}, \citenamefont {Wheeler},\ and\ \citenamefont
  {Kaiser}}]{misner2017gravitation}%
  \BibitemOpen
  \bibfield  {author} {\bibinfo {author} {\bibfnamefont {C.~W.}\ \bibnamefont
  {Misner}}, \bibinfo {author} {\bibfnamefont {K.~S.}\ \bibnamefont {Thorne}},
  \bibinfo {author} {\bibfnamefont {J.~A.}\ \bibnamefont {Wheeler}}, \ and\
  \bibinfo {author} {\bibfnamefont {D.~I.}\ \bibnamefont {Kaiser}},\
  }\href@noop {} {\emph {\bibinfo {title} {Gravitation}}}\ (\bibinfo
  {publisher} {Princeton University Press},\ \bibinfo {year}
  {2017})\BibitemShut {NoStop}%
\bibitem [{\citenamefont {Pugliese}\ \emph
  {et~al.}(2011{\natexlab{b}})\citenamefont {Pugliese}, \citenamefont
  {Quevedo},\ and\ \citenamefont {Ruffini}}]{Pugliese:2011xn}%
  \BibitemOpen
  \bibfield  {author} {\bibinfo {author} {\bibfnamefont {D.}~\bibnamefont
  {Pugliese}}, \bibinfo {author} {\bibfnamefont {H.}~\bibnamefont {Quevedo}}, \
  and\ \bibinfo {author} {\bibfnamefont {R.}~\bibnamefont {Ruffini}},\ }\href
  {\doibase 10.1103/PhysRevD.84.044030} {\bibfield  {journal} {\bibinfo
  {journal} {Phys. Rev.}\ }\textbf {\bibinfo {volume} {D84}},\ \bibinfo {pages}
  {044030} (\bibinfo {year} {2011}{\natexlab{b}})},\ \Eprint
  {http://arxiv.org/abs/1105.2959} {arXiv:1105.2959 [gr-qc]} \BibitemShut
  {NoStop}%
\bibitem [{\citenamefont {Pugliese}\ \emph
  {et~al.}(2011{\natexlab{c}})\citenamefont {Pugliese}, \citenamefont
  {Quevedo},\ and\ \citenamefont {Ruffini}}]{Pugliese:2011py}%
  \BibitemOpen
  \bibfield  {author} {\bibinfo {author} {\bibfnamefont {D.}~\bibnamefont
  {Pugliese}}, \bibinfo {author} {\bibfnamefont {H.}~\bibnamefont {Quevedo}}, \
  and\ \bibinfo {author} {\bibfnamefont {R.}~\bibnamefont {Ruffini}},\ }\href
  {\doibase 10.1103/PhysRevD.83.104052} {\bibfield  {journal} {\bibinfo
  {journal} {Phys. Rev.}\ }\textbf {\bibinfo {volume} {D83}},\ \bibinfo {pages}
  {104052} (\bibinfo {year} {2011}{\natexlab{c}})},\ \Eprint
  {http://arxiv.org/abs/1103.1807} {arXiv:1103.1807 [gr-qc]} \BibitemShut
  {NoStop}%
\bibitem [{\citenamefont {Narayan}\ and\ \citenamefont
  {McClintock}(2012{\natexlab{a}})}]{narayan2012observational}%
  \BibitemOpen
  \bibfield  {author} {\bibinfo {author} {\bibfnamefont {R.}~\bibnamefont
  {Narayan}}\ and\ \bibinfo {author} {\bibfnamefont {J.~E.}\ \bibnamefont
  {McClintock}},\ }\href@noop {} {\bibfield  {journal} {\bibinfo  {journal}
  {Monthly Notices of the Royal Astronomical Society: Letters}\ }\textbf
  {\bibinfo {volume} {419}},\ \bibinfo {pages} {L69} (\bibinfo {year}
  {2012}{\natexlab{a}})}\BibitemShut {NoStop}%
\bibitem [{\citenamefont {Tchekhovskoy}\ \emph {et~al.}(2011)\citenamefont
  {Tchekhovskoy}, \citenamefont {Narayan},\ and\ \citenamefont
  {McKinney}}]{Tchekhovskoy:2011zx}%
  \BibitemOpen
  \bibfield  {author} {\bibinfo {author} {\bibfnamefont {A.}~\bibnamefont
  {Tchekhovskoy}}, \bibinfo {author} {\bibfnamefont {R.}~\bibnamefont
  {Narayan}}, \ and\ \bibinfo {author} {\bibfnamefont {J.~C.}\ \bibnamefont
  {McKinney}},\ }\href {\doibase 10.1111/j.1745-3933.2011.01147.x} {\bibfield
  {journal} {\bibinfo  {journal} {Mon. Not. Roy. Astron. Soc.}\ }\textbf
  {\bibinfo {volume} {418}},\ \bibinfo {pages} {L79} (\bibinfo {year}
  {2011})},\ \Eprint {http://arxiv.org/abs/1108.0412} {arXiv:1108.0412
  [astro-ph.HE]} \BibitemShut {NoStop}%
\bibitem [{\citenamefont {Al~Zahrani}(2014)}]{Zahrani:2014rqa}%
  \BibitemOpen
  \bibfield  {author} {\bibinfo {author} {\bibfnamefont {A.~M.}\ \bibnamefont
  {Al~Zahrani}},\ }\href {\doibase 10.1103/PhysRevD.90.044012} {\bibfield
  {journal} {\bibinfo  {journal} {Phys. Rev.}\ }\textbf {\bibinfo {volume}
  {D90}},\ \bibinfo {pages} {044012} (\bibinfo {year} {2014})},\ \Eprint
  {http://arxiv.org/abs/1407.7069} {arXiv:1407.7069 [gr-qc]} \BibitemShut
  {NoStop}%
\bibitem [{\citenamefont {Abdujabbarov}\ and\ \citenamefont
  {Ahmedov}(2010)}]{Abdujabbarov:2009az}%
  \BibitemOpen
  \bibfield  {author} {\bibinfo {author} {\bibfnamefont {A.}~\bibnamefont
  {Abdujabbarov}}\ and\ \bibinfo {author} {\bibfnamefont {B.}~\bibnamefont
  {Ahmedov}},\ }\href {\doibase 10.1103/PhysRevD.81.044022} {\bibfield
  {journal} {\bibinfo  {journal} {Phys. Rev.}\ }\textbf {\bibinfo {volume}
  {D81}},\ \bibinfo {pages} {044022} (\bibinfo {year} {2010})},\ \Eprint
  {http://arxiv.org/abs/0905.2730} {arXiv:0905.2730 [gr-qc]} \BibitemShut
  {NoStop}%
\bibitem [{\citenamefont {Akcay}\ \emph {et~al.}(2012)\citenamefont {Akcay},
  \citenamefont {Barack}, \citenamefont {Damour},\ and\ \citenamefont
  {Sago}}]{Akcay:2012ea}%
  \BibitemOpen
  \bibfield  {author} {\bibinfo {author} {\bibfnamefont {S.}~\bibnamefont
  {Akcay}}, \bibinfo {author} {\bibfnamefont {L.}~\bibnamefont {Barack}},
  \bibinfo {author} {\bibfnamefont {T.}~\bibnamefont {Damour}}, \ and\ \bibinfo
  {author} {\bibfnamefont {N.}~\bibnamefont {Sago}},\ }\href {\doibase
  10.1103/PhysRevD.86.104041} {\bibfield  {journal} {\bibinfo  {journal} {Phys.
  Rev.}\ }\textbf {\bibinfo {volume} {D86}},\ \bibinfo {pages} {104041}
  (\bibinfo {year} {2012})},\ \Eprint {http://arxiv.org/abs/1209.0964}
  {arXiv:1209.0964 [gr-qc]} \BibitemShut {NoStop}%
\bibitem [{\citenamefont {Asano}\ and\ \citenamefont
  {Mï¿œszï¿œros}(2016)}]{Asano:2016scy}%
  \BibitemOpen
  \bibfield  {author} {\bibinfo {author} {\bibfnamefont {K.}~\bibnamefont
  {Asano}}\ and\ \bibinfo {author} {\bibfnamefont {P.}~\bibnamefont
  {Mï¿œszï¿œros}},\ }\href {\doibase 10.1103/PhysRevD.94.023005}
  {\bibfield  {journal} {\bibinfo  {journal} {Phys. Rev.}\ }\textbf {\bibinfo
  {volume} {D94}},\ \bibinfo {pages} {023005} (\bibinfo {year} {2016})},\
  \Eprint {http://arxiv.org/abs/1607.00732} {arXiv:1607.00732 [astro-ph.HE]}
  \BibitemShut {NoStop}%
\bibitem [{\citenamefont {Cabanac}\ \emph {et~al.}(2009)\citenamefont
  {Cabanac}, \citenamefont {Fender}, \citenamefont {Dunn},\ and\ \citenamefont
  {Koerding}}]{Cabanac:2009yz}%
  \BibitemOpen
  \bibfield  {author} {\bibinfo {author} {\bibfnamefont {C.}~\bibnamefont
  {Cabanac}}, \bibinfo {author} {\bibfnamefont {R.~P.}\ \bibnamefont {Fender}},
  \bibinfo {author} {\bibfnamefont {R.~J.~H.}\ \bibnamefont {Dunn}}, \ and\
  \bibinfo {author} {\bibfnamefont {E.~G.}\ \bibnamefont {Koerding}},\ }\href
  {\doibase 10.1111/j.1365-2966.2009.14867.x} {\bibfield  {journal} {\bibinfo
  {journal} {Mon. Not. Roy. Astron. Soc.}\ }\textbf {\bibinfo {volume} {396}},\
  \bibinfo {pages} {1415} (\bibinfo {year} {2009})},\ \Eprint
  {http://arxiv.org/abs/0904.0701} {arXiv:0904.0701 [astro-ph.HE]} \BibitemShut
  {NoStop}%
\bibitem [{\citenamefont {Campanelli}\ \emph {et~al.}(2006)\citenamefont
  {Campanelli}, \citenamefont {Lousto},\ and\ \citenamefont
  {Zlochower}}]{Campanelli:2006gf}%
  \BibitemOpen
  \bibfield  {author} {\bibinfo {author} {\bibfnamefont {M.}~\bibnamefont
  {Campanelli}}, \bibinfo {author} {\bibfnamefont {C.~O.}\ \bibnamefont
  {Lousto}}, \ and\ \bibinfo {author} {\bibfnamefont {Y.}~\bibnamefont
  {Zlochower}},\ }\href {\doibase 10.1103/PhysRevD.73.061501} {\bibfield
  {journal} {\bibinfo  {journal} {Phys. Rev.}\ }\textbf {\bibinfo {volume}
  {D73}},\ \bibinfo {pages} {061501} (\bibinfo {year} {2006})},\ \Eprint
  {http://arxiv.org/abs/gr-qc/0601091} {arXiv:gr-qc/0601091 [gr-qc]}
  \BibitemShut {NoStop}%
\bibitem [{\citenamefont {Chakraborty}(2014)}]{Chakraborty:2013kza}%
  \BibitemOpen
  \bibfield  {author} {\bibinfo {author} {\bibfnamefont {C.}~\bibnamefont
  {Chakraborty}},\ }\href {\doibase 10.1140/epjc/s10052-014-2759-9} {\bibfield
  {journal} {\bibinfo  {journal} {Eur. Phys. J.}\ }\textbf {\bibinfo {volume}
  {C74}},\ \bibinfo {pages} {2759} (\bibinfo {year} {2014})},\ \Eprint
  {http://arxiv.org/abs/1307.4698} {arXiv:1307.4698 [gr-qc]} \BibitemShut
  {NoStop}%
\bibitem [{\citenamefont {Delsate}\ \emph {et~al.}(2015)\citenamefont
  {Delsate}, \citenamefont {Rocha},\ and\ \citenamefont
  {Santarelli}}]{Delsate:2015ina}%
  \BibitemOpen
  \bibfield  {author} {\bibinfo {author} {\bibfnamefont {T.}~\bibnamefont
  {Delsate}}, \bibinfo {author} {\bibfnamefont {J.~V.}\ \bibnamefont {Rocha}},
  \ and\ \bibinfo {author} {\bibfnamefont {R.}~\bibnamefont {Santarelli}},\
  }\href {\doibase 10.1103/PhysRevD.92.084028} {\bibfield  {journal} {\bibinfo
  {journal} {Phys. Rev.}\ }\textbf {\bibinfo {volume} {D92}},\ \bibinfo {pages}
  {084028} (\bibinfo {year} {2015})},\ \Eprint
  {http://arxiv.org/abs/1507.03602} {arXiv:1507.03602 [gr-qc]} \BibitemShut
  {NoStop}%
\bibitem [{\citenamefont {Hadar}\ \emph {et~al.}(2011)\citenamefont {Hadar},
  \citenamefont {Kol}, \citenamefont {Berti},\ and\ \citenamefont
  {Cardoso}}]{Hadar:2011vj}%
  \BibitemOpen
  \bibfield  {author} {\bibinfo {author} {\bibfnamefont {S.}~\bibnamefont
  {Hadar}}, \bibinfo {author} {\bibfnamefont {B.}~\bibnamefont {Kol}}, \bibinfo
  {author} {\bibfnamefont {E.}~\bibnamefont {Berti}}, \ and\ \bibinfo {author}
  {\bibfnamefont {V.}~\bibnamefont {Cardoso}},\ }\href {\doibase
  10.1103/PhysRevD.84.047501} {\bibfield  {journal} {\bibinfo  {journal} {Phys.
  Rev.}\ }\textbf {\bibinfo {volume} {D84}},\ \bibinfo {pages} {047501}
  (\bibinfo {year} {2011})},\ \Eprint {http://arxiv.org/abs/1105.3861}
  {arXiv:1105.3861 [gr-qc]} \BibitemShut {NoStop}%
\bibitem [{\citenamefont {Harms}\ \emph
  {et~al.}(2016{\natexlab{a}})\citenamefont {Harms}, \citenamefont
  {Lukes-Gerakopoulos}, \citenamefont {Bernuzzi},\ and\ \citenamefont
  {Nagar}}]{Harms:2016ctx}%
  \BibitemOpen
  \bibfield  {author} {\bibinfo {author} {\bibfnamefont {E.}~\bibnamefont
  {Harms}}, \bibinfo {author} {\bibfnamefont {G.}~\bibnamefont
  {Lukes-Gerakopoulos}}, \bibinfo {author} {\bibfnamefont {S.}~\bibnamefont
  {Bernuzzi}}, \ and\ \bibinfo {author} {\bibfnamefont {A.}~\bibnamefont
  {Nagar}},\ }\href {\doibase 10.1103/PhysRevD.94.104010} {\bibfield  {journal}
  {\bibinfo  {journal} {Phys. Rev.}\ }\textbf {\bibinfo {volume} {D94}},\
  \bibinfo {pages} {104010} (\bibinfo {year} {2016}{\natexlab{a}})},\ \Eprint
  {http://arxiv.org/abs/1609.00356} {arXiv:1609.00356 [gr-qc]} \BibitemShut
  {NoStop}%
\bibitem [{\citenamefont {Hod}(2013)}]{Hod:2013vi}%
  \BibitemOpen
  \bibfield  {author} {\bibinfo {author} {\bibfnamefont {S.}~\bibnamefont
  {Hod}},\ }\href {\doibase 10.1103/PhysRevD.87.024036} {\bibfield  {journal}
  {\bibinfo  {journal} {Phys. Rev.}\ }\textbf {\bibinfo {volume} {D87}},\
  \bibinfo {pages} {024036} (\bibinfo {year} {2013})},\ \Eprint
  {http://arxiv.org/abs/1311.1281} {arXiv:1311.1281 [gr-qc]} \BibitemShut
  {NoStop}%
\bibitem [{\citenamefont {Hod}(2014)}]{Hod:2014tpa}%
  \BibitemOpen
  \bibfield  {author} {\bibinfo {author} {\bibfnamefont {S.}~\bibnamefont
  {Hod}},\ }\href {\doibase 10.1140/epjc/s10052-014-2840-4} {\bibfield
  {journal} {\bibinfo  {journal} {Eur. Phys. J.}\ }\textbf {\bibinfo {volume}
  {C74}},\ \bibinfo {pages} {2840} (\bibinfo {year} {2014})},\ \Eprint
  {http://arxiv.org/abs/1404.1566} {arXiv:1404.1566 [gr-qc]} \BibitemShut
  {NoStop}%
\bibitem [{\citenamefont {Isoyama}\ \emph {et~al.}(2014)\citenamefont
  {Isoyama}, \citenamefont {Barack}, \citenamefont {Dolan}, \citenamefont
  {Le~Tiec}, \citenamefont {Nakano}, \citenamefont {Shah}, \citenamefont
  {Tanaka},\ and\ \citenamefont {Warburton}}]{Isoyama:2014mja}%
  \BibitemOpen
  \bibfield  {author} {\bibinfo {author} {\bibfnamefont {S.}~\bibnamefont
  {Isoyama}}, \bibinfo {author} {\bibfnamefont {L.}~\bibnamefont {Barack}},
  \bibinfo {author} {\bibfnamefont {S.~R.}\ \bibnamefont {Dolan}}, \bibinfo
  {author} {\bibfnamefont {A.}~\bibnamefont {Le~Tiec}}, \bibinfo {author}
  {\bibfnamefont {H.}~\bibnamefont {Nakano}}, \bibinfo {author} {\bibfnamefont
  {A.~G.}\ \bibnamefont {Shah}}, \bibinfo {author} {\bibfnamefont
  {T.}~\bibnamefont {Tanaka}}, \ and\ \bibinfo {author} {\bibfnamefont
  {N.}~\bibnamefont {Warburton}},\ }\href {\doibase
  10.1103/PhysRevLett.113.161101} {\bibfield  {journal} {\bibinfo  {journal}
  {Phys. Rev. Lett.}\ }\textbf {\bibinfo {volume} {113}},\ \bibinfo {pages}
  {161101} (\bibinfo {year} {2014})},\ \Eprint {http://arxiv.org/abs/1404.6133}
  {arXiv:1404.6133 [gr-qc]} \BibitemShut {NoStop}%
\bibitem [{\citenamefont {Lukes-Gerakopoulos}\ \emph
  {et~al.}(2017)\citenamefont {Lukes-Gerakopoulos}, \citenamefont {Harms},
  \citenamefont {Bernuzzi},\ and\ \citenamefont
  {Nagar}}]{Lukes-Gerakopoulos:2017vkj}%
  \BibitemOpen
  \bibfield  {author} {\bibinfo {author} {\bibfnamefont {G.}~\bibnamefont
  {Lukes-Gerakopoulos}}, \bibinfo {author} {\bibfnamefont {E.}~\bibnamefont
  {Harms}}, \bibinfo {author} {\bibfnamefont {S.}~\bibnamefont {Bernuzzi}}, \
  and\ \bibinfo {author} {\bibfnamefont {A.}~\bibnamefont {Nagar}},\ }\href
  {\doibase 10.1103/PhysRevD.96.064051} {\bibfield  {journal} {\bibinfo
  {journal} {Phys. Rev.}\ }\textbf {\bibinfo {volume} {D96}},\ \bibinfo {pages}
  {064051} (\bibinfo {year} {2017})},\ \Eprint
  {http://arxiv.org/abs/1707.07537} {arXiv:1707.07537 [gr-qc]} \BibitemShut
  {NoStop}%
\bibitem [{\citenamefont {Zaslavskii}(2015)}]{Zaslavskii:2014mqa}%
  \BibitemOpen
  \bibfield  {author} {\bibinfo {author} {\bibfnamefont {O.~B.}\ \bibnamefont
  {Zaslavskii}},\ }\href {\doibase 10.1140/epjc/s10052-015-3623-2} {\bibfield
  {journal} {\bibinfo  {journal} {Eur. Phys. J.}\ }\textbf {\bibinfo {volume}
  {C75}},\ \bibinfo {pages} {403} (\bibinfo {year} {2015})},\ \Eprint
  {http://arxiv.org/abs/1405.2543} {arXiv:1405.2543 [gr-qc]} \BibitemShut
  {NoStop}%
\bibitem [{\citenamefont {Shiose}\ \emph {et~al.}(2014)\citenamefont {Shiose},
  \citenamefont {Kimura},\ and\ \citenamefont {Chiba}}]{Shiose:2014bqa}%
  \BibitemOpen
  \bibfield  {author} {\bibinfo {author} {\bibfnamefont {R.}~\bibnamefont
  {Shiose}}, \bibinfo {author} {\bibfnamefont {M.}~\bibnamefont {Kimura}}, \
  and\ \bibinfo {author} {\bibfnamefont {T.}~\bibnamefont {Chiba}},\ }\href
  {\doibase 10.1103/PhysRevD.90.124016} {\bibfield  {journal} {\bibinfo
  {journal} {Phys. Rev.}\ }\textbf {\bibinfo {volume} {D90}},\ \bibinfo {pages}
  {124016} (\bibinfo {year} {2014})},\ \Eprint {http://arxiv.org/abs/1409.3310}
  {arXiv:1409.3310 [gr-qc]} \BibitemShut {NoStop}%
\bibitem [{\citenamefont {Hussain}\ \emph {et~al.}(2014)\citenamefont
  {Hussain}, \citenamefont {Hussain},\ and\ \citenamefont
  {Jamil}}]{Hussain:2014cba}%
  \BibitemOpen
  \bibfield  {author} {\bibinfo {author} {\bibfnamefont {S.}~\bibnamefont
  {Hussain}}, \bibinfo {author} {\bibfnamefont {I.}~\bibnamefont {Hussain}}, \
  and\ \bibinfo {author} {\bibfnamefont {M.}~\bibnamefont {Jamil}},\ }\href
  {\doibase 10.1140/epjc/s10052-014-3210-y} {\bibfield  {journal} {\bibinfo
  {journal} {Eur. Phys. J.}\ }\textbf {\bibinfo {volume} {C74}},\ \bibinfo
  {pages} {3210} (\bibinfo {year} {2014})},\ \Eprint
  {http://arxiv.org/abs/1402.2731} {arXiv:1402.2731 [gr-qc]} \BibitemShut
  {NoStop}%
\bibitem [{\citenamefont {Takahashi}\ and\ \citenamefont
  {Koyama}(2009)}]{Takahashi:2008zh}%
  \BibitemOpen
  \bibfield  {author} {\bibinfo {author} {\bibfnamefont {M.}~\bibnamefont
  {Takahashi}}\ and\ \bibinfo {author} {\bibfnamefont {H.}~\bibnamefont
  {Koyama}},\ }\href {\doibase 10.1088/0004-637X/693/1/472} {\bibfield
  {journal} {\bibinfo  {journal} {Astrophys. J.}\ }\textbf {\bibinfo {volume}
  {693}},\ \bibinfo {pages} {472} (\bibinfo {year} {2009})},\ \Eprint
  {http://arxiv.org/abs/0807.0277} {arXiv:0807.0277 [astro-ph]} \BibitemShut
  {NoStop}%
\bibitem [{\citenamefont {Frolov}\ and\ \citenamefont
  {Shoom}(2010)}]{Frolov:2010mi}%
  \BibitemOpen
  \bibfield  {author} {\bibinfo {author} {\bibfnamefont {V.~P.}\ \bibnamefont
  {Frolov}}\ and\ \bibinfo {author} {\bibfnamefont {A.~A.}\ \bibnamefont
  {Shoom}},\ }\href {\doibase 10.1103/PhysRevD.82.084034} {\bibfield  {journal}
  {\bibinfo  {journal} {Phys. Rev.}\ }\textbf {\bibinfo {volume} {D82}},\
  \bibinfo {pages} {084034} (\bibinfo {year} {2010})},\ \Eprint
  {http://arxiv.org/abs/1008.2985} {arXiv:1008.2985 [gr-qc]} \BibitemShut
  {NoStop}%
\bibitem [{\citenamefont {Frolov}(2012)}]{Frolov:2011ea}%
  \BibitemOpen
  \bibfield  {author} {\bibinfo {author} {\bibfnamefont {V.~P.}\ \bibnamefont
  {Frolov}},\ }\href {\doibase 10.1103/PhysRevD.85.024020} {\bibfield
  {journal} {\bibinfo  {journal} {Phys. Rev.}\ }\textbf {\bibinfo {volume}
  {D85}},\ \bibinfo {pages} {024020} (\bibinfo {year} {2012})},\ \Eprint
  {http://arxiv.org/abs/1110.6274} {arXiv:1110.6274 [gr-qc]} \BibitemShut
  {NoStop}%
\bibitem [{\citenamefont {Pugliese}\ \emph {et~al.}(2013)\citenamefont
  {Pugliese}, \citenamefont {Quevedo},\ and\ \citenamefont
  {Ruffini}}]{Pugliese:2013zma}%
  \BibitemOpen
  \bibfield  {author} {\bibinfo {author} {\bibfnamefont {D.}~\bibnamefont
  {Pugliese}}, \bibinfo {author} {\bibfnamefont {H.}~\bibnamefont {Quevedo}}, \
  and\ \bibinfo {author} {\bibfnamefont {R.}~\bibnamefont {Ruffini}},\ }\href
  {\doibase 10.1103/PhysRevD.88.024042} {\bibfield  {journal} {\bibinfo
  {journal} {Phys. Rev.}\ }\textbf {\bibinfo {volume} {D88}},\ \bibinfo {pages}
  {024042} (\bibinfo {year} {2013})},\ \Eprint {http://arxiv.org/abs/1303.6250}
  {arXiv:1303.6250 [gr-qc]} \BibitemShut {NoStop}%
\bibitem [{\citenamefont {Zhang}\ \emph
  {et~al.}(2018{\natexlab{a}})\citenamefont {Zhang}, \citenamefont {Wei},
  \citenamefont {Amaro-Seoane}, \citenamefont {Yang},\ and\ \citenamefont
  {Liu}}]{Zhang:2018omr}%
  \BibitemOpen
  \bibfield  {author} {\bibinfo {author} {\bibfnamefont {Y.-P.}\ \bibnamefont
  {Zhang}}, \bibinfo {author} {\bibfnamefont {S.-W.}\ \bibnamefont {Wei}},
  \bibinfo {author} {\bibfnamefont {P.}~\bibnamefont {Amaro-Seoane}}, \bibinfo
  {author} {\bibfnamefont {J.}~\bibnamefont {Yang}}, \ and\ \bibinfo {author}
  {\bibfnamefont {Y.-X.}\ \bibnamefont {Liu}},\ }\href@noop {} {\  (\bibinfo
  {year} {2018}{\natexlab{a}})},\ \Eprint {http://arxiv.org/abs/1812.06345}
  {arXiv:1812.06345 [gr-qc]} \BibitemShut {NoStop}%
\bibitem [{\citenamefont {Schroven}\ \emph {et~al.}(2017)\citenamefont
  {Schroven}, \citenamefont {Hackmann},\ and\ \citenamefont
  {Lï¿œmmerzahl}}]{Schroven:2017jsp}%
  \BibitemOpen
  \bibfield  {author} {\bibinfo {author} {\bibfnamefont {K.}~\bibnamefont
  {Schroven}}, \bibinfo {author} {\bibfnamefont {E.}~\bibnamefont {Hackmann}},
  \ and\ \bibinfo {author} {\bibfnamefont {C.}~\bibnamefont
  {Lï¿œmmerzahl}},\ }\href {\doibase 10.1103/PhysRevD.96.063015} {\bibfield
  {journal} {\bibinfo  {journal} {Phys. Rev.}\ }\textbf {\bibinfo {volume}
  {D96}},\ \bibinfo {pages} {063015} (\bibinfo {year} {2017})},\ \Eprint
  {http://arxiv.org/abs/1705.08166} {arXiv:1705.08166 [astro-ph.HE]}
  \BibitemShut {NoStop}%
\bibitem [{\citenamefont {Corinaldesi}\ and\ \citenamefont
  {Papapetrou}(1951)}]{Corinaldesi:1951pb}%
  \BibitemOpen
  \bibfield  {author} {\bibinfo {author} {\bibfnamefont {E.}~\bibnamefont
  {Corinaldesi}}\ and\ \bibinfo {author} {\bibfnamefont {A.}~\bibnamefont
  {Papapetrou}},\ }\href {\doibase 10.1098/rspa.1951.0201} {\bibfield
  {journal} {\bibinfo  {journal} {Proc. Roy. Soc. Lond.}\ }\textbf {\bibinfo
  {volume} {A209}},\ \bibinfo {pages} {259} (\bibinfo {year}
  {1951})}\BibitemShut {NoStop}%
\bibitem [{\citenamefont {Rasband}(1973)}]{rasband1973black}%
  \BibitemOpen
  \bibfield  {author} {\bibinfo {author} {\bibfnamefont {S.}~\bibnamefont
  {Rasband}},\ }\href@noop {} {\bibfield  {journal} {\bibinfo  {journal} {Phys.
  Rev. Lett.}\ }\textbf {\bibinfo {volume} {30}},\ \bibinfo {pages} {111}
  (\bibinfo {year} {1973})}\BibitemShut {NoStop}%
\bibitem [{\citenamefont {Tod}\ \emph {et~al.}(1976)\citenamefont {Tod},
  \citenamefont {de~Felice},\ and\ \citenamefont {Calvani}}]{Tod:1976ud}%
  \BibitemOpen
  \bibfield  {author} {\bibinfo {author} {\bibfnamefont {K.~P.}\ \bibnamefont
  {Tod}}, \bibinfo {author} {\bibfnamefont {F.}~\bibnamefont {de~Felice}}, \
  and\ \bibinfo {author} {\bibfnamefont {M.}~\bibnamefont {Calvani}},\ }\href
  {\doibase 10.1007/BF02728614} {\bibfield  {journal} {\bibinfo  {journal}
  {Nuovo Cim.}\ }\textbf {\bibinfo {volume} {B34}},\ \bibinfo {pages} {365}
  (\bibinfo {year} {1976})}\BibitemShut {NoStop}%
\bibitem [{\citenamefont {Zhang}\ \emph
  {et~al.}(2018{\natexlab{b}})\citenamefont {Zhang}, \citenamefont {Wei},
  \citenamefont {Guo}, \citenamefont {Sui},\ and\ \citenamefont
  {Liu}}]{Zhang:2017nhl}%
  \BibitemOpen
  \bibfield  {author} {\bibinfo {author} {\bibfnamefont {Y.-P.}\ \bibnamefont
  {Zhang}}, \bibinfo {author} {\bibfnamefont {S.-W.}\ \bibnamefont {Wei}},
  \bibinfo {author} {\bibfnamefont {W.-D.}\ \bibnamefont {Guo}}, \bibinfo
  {author} {\bibfnamefont {T.-T.}\ \bibnamefont {Sui}}, \ and\ \bibinfo
  {author} {\bibfnamefont {Y.-X.}\ \bibnamefont {Liu}},\ }\href {\doibase
  10.1103/PhysRevD.97.084056} {\bibfield  {journal} {\bibinfo  {journal} {Phys.
  Rev.}\ }\textbf {\bibinfo {volume} {D97}},\ \bibinfo {pages} {084056}
  (\bibinfo {year} {2018}{\natexlab{b}})},\ \Eprint
  {http://arxiv.org/abs/1711.09361} {arXiv:1711.09361 [gr-qc]} \BibitemShut
  {NoStop}%
\bibitem [{\citenamefont {Hojman}\ and\ \citenamefont
  {Hojman}(1977)}]{Hojman:1976kn}%
  \BibitemOpen
  \bibfield  {author} {\bibinfo {author} {\bibfnamefont {R.}~\bibnamefont
  {Hojman}}\ and\ \bibinfo {author} {\bibfnamefont {S.}~\bibnamefont
  {Hojman}},\ }\href {\doibase 10.1103/PhysRevD.15.2724} {\bibfield  {journal}
  {\bibinfo  {journal} {Phys. Rev.}\ }\textbf {\bibinfo {volume} {D15}},\
  \bibinfo {pages} {2724} (\bibinfo {year} {1977})}\BibitemShut {NoStop}%
\bibitem [{\citenamefont {Saijo}\ \emph {et~al.}(1998)\citenamefont {Saijo},
  \citenamefont {Maeda}, \citenamefont {Shibata},\ and\ \citenamefont
  {Mino}}]{Saijo:1998mn}%
  \BibitemOpen
  \bibfield  {author} {\bibinfo {author} {\bibfnamefont {M.}~\bibnamefont
  {Saijo}}, \bibinfo {author} {\bibfnamefont {K.-i.}\ \bibnamefont {Maeda}},
  \bibinfo {author} {\bibfnamefont {M.}~\bibnamefont {Shibata}}, \ and\
  \bibinfo {author} {\bibfnamefont {Y.}~\bibnamefont {Mino}},\ }\href {\doibase
  10.1103/PhysRevD.58.064005} {\bibfield  {journal} {\bibinfo  {journal} {Phys.
  Rev.}\ }\textbf {\bibinfo {volume} {D58}},\ \bibinfo {pages} {064005}
  (\bibinfo {year} {1998})}\BibitemShut {NoStop}%
\bibitem [{\citenamefont {Wald}(1972)}]{Wald:1972sz}%
  \BibitemOpen
  \bibfield  {author} {\bibinfo {author} {\bibfnamefont {R.~M.}\ \bibnamefont
  {Wald}},\ }\href {\doibase 10.1103/PhysRevD.6.406} {\bibfield  {journal}
  {\bibinfo  {journal} {Phys. Rev.}\ }\textbf {\bibinfo {volume} {D6}},\
  \bibinfo {pages} {406} (\bibinfo {year} {1972})}\BibitemShut {NoStop}%
\bibitem [{\citenamefont {Narayan}\ and\ \citenamefont
  {McClintock}(2012{\natexlab{b}})}]{Narayan:2011eb}%
  \BibitemOpen
  \bibfield  {author} {\bibinfo {author} {\bibfnamefont {R.}~\bibnamefont
  {Narayan}}\ and\ \bibinfo {author} {\bibfnamefont {J.~E.}\ \bibnamefont
  {McClintock}},\ }\href {\doibase 10.1111/j.1745-3933.2011.01181.x} {\bibfield
   {journal} {\bibinfo  {journal} {Mon. Not. Roy. Astron. Soc.}\ }\textbf
  {\bibinfo {volume} {419}},\ \bibinfo {pages} {L69} (\bibinfo {year}
  {2012}{\natexlab{b}})},\ \Eprint {http://arxiv.org/abs/1112.0569}
  {arXiv:1112.0569 [astro-ph.HE]} \BibitemShut {NoStop}%
\bibitem [{\citenamefont {Zhang}\ \emph
  {et~al.}(2018{\natexlab{c}})\citenamefont {Zhang}, \citenamefont {Jiang},
  \citenamefont {Liu},\ and\ \citenamefont {Liu}}]{Zhang:2018gpn}%
  \BibitemOpen
  \bibfield  {author} {\bibinfo {author} {\bibfnamefont {M.}~\bibnamefont
  {Zhang}}, \bibinfo {author} {\bibfnamefont {J.}~\bibnamefont {Jiang}},
  \bibinfo {author} {\bibfnamefont {Y.}~\bibnamefont {Liu}}, \ and\ \bibinfo
  {author} {\bibfnamefont {W.-B.}\ \bibnamefont {Liu}},\ }\href {\doibase
  10.1103/PhysRevD.98.044006} {\bibfield  {journal} {\bibinfo  {journal} {Phys.
  Rev.}\ }\textbf {\bibinfo {volume} {D98}},\ \bibinfo {pages} {044006}
  (\bibinfo {year} {2018}{\natexlab{c}})}\BibitemShut {NoStop}%
\bibitem [{\citenamefont {Castineiras}\ \emph {et~al.}(2005)\citenamefont
  {Castineiras}, \citenamefont {Crispino}, \citenamefont {Murta},\ and\
  \citenamefont {Matsas}}]{Castineiras:2005ww}%
  \BibitemOpen
  \bibfield  {author} {\bibinfo {author} {\bibfnamefont {J.}~\bibnamefont
  {Castineiras}}, \bibinfo {author} {\bibfnamefont {L.~C.~B.}\ \bibnamefont
  {Crispino}}, \bibinfo {author} {\bibfnamefont {R.}~\bibnamefont {Murta}}, \
  and\ \bibinfo {author} {\bibfnamefont {G.~E.~A.}\ \bibnamefont {Matsas}},\
  }\href {\doibase 10.1103/PhysRevD.71.104013} {\bibfield  {journal} {\bibinfo
  {journal} {Phys. Rev.}\ }\textbf {\bibinfo {volume} {D71}},\ \bibinfo {pages}
  {104013} (\bibinfo {year} {2005})},\ \Eprint
  {http://arxiv.org/abs/gr-qc/0503050} {arXiv:gr-qc/0503050 [gr-qc]}
  \BibitemShut {NoStop}%
\bibitem [{\citenamefont {Misner}(1972)}]{Misner:1972kx}%
  \BibitemOpen
  \bibfield  {author} {\bibinfo {author} {\bibfnamefont {C.~W.}\ \bibnamefont
  {Misner}},\ }\href {\doibase 10.1103/PhysRevLett.28.994} {\bibfield
  {journal} {\bibinfo  {journal} {Phys. Rev. Lett.}\ }\textbf {\bibinfo
  {volume} {28}},\ \bibinfo {pages} {994} (\bibinfo {year} {1972})}\BibitemShut
  {NoStop}%
\bibitem [{\citenamefont {Misner}\ \emph {et~al.}(1972)\citenamefont {Misner},
  \citenamefont {Breuer}, \citenamefont {Brill}, \citenamefont {Chrzanowski},
  \citenamefont {Hughes},\ and\ \citenamefont {Pereira}}]{Misner:1972jf}%
  \BibitemOpen
  \bibfield  {author} {\bibinfo {author} {\bibfnamefont {C.~W.}\ \bibnamefont
  {Misner}}, \bibinfo {author} {\bibfnamefont {R.~A.}\ \bibnamefont {Breuer}},
  \bibinfo {author} {\bibfnamefont {D.~R.}\ \bibnamefont {Brill}}, \bibinfo
  {author} {\bibfnamefont {P.~L.}\ \bibnamefont {Chrzanowski}}, \bibinfo
  {author} {\bibfnamefont {H.~G.}\ \bibnamefont {Hughes}}, \ and\ \bibinfo
  {author} {\bibfnamefont {C.~M.}\ \bibnamefont {Pereira}},\ }\href {\doibase
  10.1103/PhysRevLett.28.998} {\bibfield  {journal} {\bibinfo  {journal} {Phys.
  Rev. Lett.}\ }\textbf {\bibinfo {volume} {28}},\ \bibinfo {pages} {998}
  (\bibinfo {year} {1972})}\BibitemShut {NoStop}%
\bibitem [{\citenamefont {Poisson}(1993)}]{Poisson:1993vp}%
  \BibitemOpen
  \bibfield  {author} {\bibinfo {author} {\bibfnamefont {E.}~\bibnamefont
  {Poisson}},\ }\href {\doibase 10.1103/PhysRevD.47.1497} {\bibfield  {journal}
  {\bibinfo  {journal} {Phys. Rev.}\ }\textbf {\bibinfo {volume} {D47}},\
  \bibinfo {pages} {1497} (\bibinfo {year} {1993})}\BibitemShut {NoStop}%
\bibitem [{\citenamefont {Cutler}\ \emph {et~al.}(1993)\citenamefont {Cutler},
  \citenamefont {Poisson}, \citenamefont {Sussman},\ and\ \citenamefont
  {Finn}}]{Cutler:1993vq}%
  \BibitemOpen
  \bibfield  {author} {\bibinfo {author} {\bibfnamefont {C.}~\bibnamefont
  {Cutler}}, \bibinfo {author} {\bibfnamefont {E.}~\bibnamefont {Poisson}},
  \bibinfo {author} {\bibfnamefont {G.~J.}\ \bibnamefont {Sussman}}, \ and\
  \bibinfo {author} {\bibfnamefont {L.~S.}\ \bibnamefont {Finn}},\ }\href
  {\doibase 10.1103/PhysRevD.47.1511} {\bibfield  {journal} {\bibinfo
  {journal} {Phys. Rev.}\ }\textbf {\bibinfo {volume} {D47}},\ \bibinfo {pages}
  {1511} (\bibinfo {year} {1993})}\BibitemShut {NoStop}%
\bibitem [{\citenamefont {Bernar}\ \emph {et~al.}(2017)\citenamefont {Bernar},
  \citenamefont {Crispino},\ and\ \citenamefont {Higuchi}}]{Bernar:2017kug}%
  \BibitemOpen
  \bibfield  {author} {\bibinfo {author} {\bibfnamefont {R.~P.}\ \bibnamefont
  {Bernar}}, \bibinfo {author} {\bibfnamefont {L.~C.}\ \bibnamefont
  {Crispino}}, \ and\ \bibinfo {author} {\bibfnamefont {A.}~\bibnamefont
  {Higuchi}},\ }\href {\doibase 10.1103/PhysRevD.95.064042} {\bibfield
  {journal} {\bibinfo  {journal} {Phys. Rev.}\ }\textbf {\bibinfo {volume}
  {D95}},\ \bibinfo {pages} {064042} (\bibinfo {year} {2017})},\ \Eprint
  {http://arxiv.org/abs/1703.10648} {arXiv:1703.10648 [gr-qc]} \BibitemShut
  {NoStop}%
\bibitem [{\citenamefont {Han}(2010)}]{Han:2010tp}%
  \BibitemOpen
  \bibfield  {author} {\bibinfo {author} {\bibfnamefont {W.-B.}\ \bibnamefont
  {Han}},\ }\href {\doibase 10.1103/PhysRevD.82.084013} {\bibfield  {journal}
  {\bibinfo  {journal} {Phys. Rev.}\ }\textbf {\bibinfo {volume} {D82}},\
  \bibinfo {pages} {084013} (\bibinfo {year} {2010})},\ \Eprint
  {http://arxiv.org/abs/1008.3324} {arXiv:1008.3324 [gr-qc]} \BibitemShut
  {NoStop}%
\bibitem [{\citenamefont {Harms}\ \emph
  {et~al.}(2016{\natexlab{b}})\citenamefont {Harms}, \citenamefont
  {Lukes-Gerakopoulos}, \citenamefont {Bernuzzi},\ and\ \citenamefont
  {Nagar}}]{Harms:2015ixa}%
  \BibitemOpen
  \bibfield  {author} {\bibinfo {author} {\bibfnamefont {E.}~\bibnamefont
  {Harms}}, \bibinfo {author} {\bibfnamefont {G.}~\bibnamefont
  {Lukes-Gerakopoulos}}, \bibinfo {author} {\bibfnamefont {S.}~\bibnamefont
  {Bernuzzi}}, \ and\ \bibinfo {author} {\bibfnamefont {A.}~\bibnamefont
  {Nagar}},\ }\href {\doibase 10.1103/PhysRevD.93.044015} {\bibfield  {journal}
  {\bibinfo  {journal} {Phys. Rev.}\ }\textbf {\bibinfo {volume} {D93}},\
  \bibinfo {pages} {044015} (\bibinfo {year} {2016}{\natexlab{b}})},\ \Eprint
  {http://arxiv.org/abs/1510.05548} {arXiv:1510.05548 [gr-qc]} \BibitemShut
  {NoStop}%
\end{thebibliography}%

\end{document}